\documentclass[sigconf]{acmart}

\usepackage{amsmath,amssymb,amsfonts}
\usepackage{algorithmic}
\usepackage{graphicx}
\usepackage{textcomp}
\usepackage{booktabs}
\usepackage{multirow}
\usepackage{bm}
\usepackage{balance}
\usepackage{float}
\usepackage{enumerate}

\usepackage{framed}

\usepackage{xcolor}

\AtBeginDocument{%
  \providecommand\BibTeX{{%
    \normalfont B\kern-0.5em{\scshape i\kern-0.25em b}\kern-0.8em\TeX}}}

\copyrightyear{2022}
\acmYear{2022}
\setcopyright{acmcopyright}\acmConference[ESEM '22]{ACM / IEEE International Symposium on Empirical Software Engineering and Measurement (ESEM)}{September 19--23, 2022}{Helsinki, Finland}
\acmBooktitle{ACM / IEEE International Symposium on Empirical Software Engineering and Measurement (ESEM) (ESEM '22), September 19--23, 2022, Helsinki, Finland}
\acmPrice{15.00}
\acmDOI{10.1145/3544902.3546253}
\acmISBN{978-1-4503-9427-7/22/09}

\begin{document}

\title[Potential Technical Debt and Its Resolution in Code Reviews]{Potential Technical Debt and Its Resolution in Code Reviews: An Exploratory Study of the OpenStack and Qt Communities}


\author{Liming Fu}
\affiliation{
\institution{School of Computer Science\\Wuhan University}
\city{Wuhan}
\country{China}
}
\email{limingfu@whu.edu.cn}

\author{Peng Liang$^{*}$}
\affiliation{
\institution{School of Computer Science\\Wuhan University}
\city{Wuhan}
\country{China}
}
\email{liangp@whu.edu.cn}

\author{Zeeshan Rasheed}
\affiliation{
\institution{School of Computer Science\\Wuhan University}
\city{Wuhan}
\country{China}
}
\email{zeeshanrasheed@whu.edu.cn}

\author{Zengyang Li}
\affiliation{
\institution{School of Computer Science\\Central China Normal University}
\city{Wuhan}
\country{China}
}
\email{zengyangli@ccnu.edu.cn}

\author{Amjed Tahir}
\affiliation{
\institution{School of Mathematical and Computational Sciences\\Massey University}
\city{Palmerston North}
\country{New Zealand}
}
\email{a.tahir@massey.ac.nz}

\author{Xiaofeng Han}
\affiliation{
\institution{School of Computer Science\\Wuhan University}
\city{Wuhan}
\country{China}
}
\email{hanxiaofeng@whu.edu.cn}


\renewcommand{\shortauthors}{L. Fu et al.}

\begin{abstract}
\textbf{Background:} Technical Debt (TD) refers to the situation where developers make trade-offs to achieve short-term goals at the expense of long-term code quality, which can have a negative impact on the quality of software systems. In the context of code review, such sub-optimal implementations have chances to be timely resolved during the review process before the code is merged. Therefore, we could consider them as Potential Technical Debt (PTD) since PTD will evolve into TD when it is injected into software systems without being resolved. \textbf{Aim:} To date, little is known about the extent to which PTD is identified in code reviews. Many tools have been provided to detect TD, but these tools lack consensus and a large amount of PTD are undetectable by tools while code review could help verify the quality of code that has been committed by identifying issues, such as PTD. To this end, we conducted an exploratory study in an attempt to understand the nature of PTD in code reviews and track down the resolution of PTD after being identified. \textbf{Method:} We randomly collected 2,030 review comments from the Nova project of OpenStack and the Qt Base project of Qt. We then manually checked these review comments, and obtained 163 PTD-related review comments for further analysis. \textbf{Results:} Our results show that: (1) PTD can be identified in code reviews but is not prevalent. (2) Design, defect, documentation, requirement, test, and code PTD are identified in code reviews, in which code and documentation PTD are the dominant. (3) 81.0\% of the PTD identified in code reviews has been resolved by developers, and 78.0\% of the resolved TD was resolved by developers within a week. (4) Code refactoring is the main practice used by developers to resolve the PTD identified in code reviews. \textbf{Conclusions:} Our findings indicate that: (1) review-based detection of PTD is seen as one of the trustworthy mechanisms in development, and (2) there is still a significant proportion of PTD (19.0\%) remaining unresolved when injected into the software systems. Practitioners and researchers should establish effective strategies to manage and resolve PTD in development.
\end{abstract}

\ccsdesc[500]{Software and its engineering~Designing software}
\ccsdesc[500]{General and reference~Empirical studies}
\keywords{Technical Debt, Potential Technical Debt, Code Review, OpenStack, Qt}
\maketitle

\section{Introduction} \label{Introduction}
Developers sometimes make compromised decisions between a shorter completion time and better software quality, and might take shortcuts to meet short-term goals, such as the need for releasing a new feature or fixing a bug. These technical compromises can yield quick and short-term benefits, but may also hurt the long-term health of a software system \cite{li2015systematic}. In this context, the Technical Debt (TD) metaphor was framed by Cunningham \cite{CunninghamMS}, which refers to the unavoidable maintenance and evolution costs of such not-quite-right solutions in development. Prior studies have shown that TD can have a negative impact on the quality of software systems \cite{DesignDebtZazworka}. 

Over the past two decades, many studies have focused on TD and contributed to the community about the nature, detection, and repayment of TD, especially Self-Admitted Technical Debt (SATD), which refers to the situation where developers know that the current implementation is not optimal and write comments alerting the inadequacy of the solution \cite{MaldonadoDetect}. However, to the best of our knowledge, little is known about the presence of TD in code reviews. Code review is the process of manually reviewing and verifying the code that has been committed by developers. It can verify a wide range of issues in the code from potential bugs and security violations to code quality issues (e.g., code smells). Those issues can also include TD because developers may intentionally or unintentionally write code that can be considered as sub-optimal during the development process \cite{flower}. Fortunately, after code reviewers detect the issues, such sub-optimal implementations have the chances to be timely resolved or improved before the code is merged into the main code repository. In this context, those issues can also be considered as Potential Technical Debt (PTD). 

The definition of PTD was first proposed by Schmid \cite{6608681}, which describe it as a risk that can potentially become a problem. In this work, we adopted the definition of PTD as the early stage of TD in code reviews, because PTD will evolve into TD if it has been introduced into software systems without being resolved, otherwise it would not have any real impact on the quality of the software. Considering the fact that PTD is a major source of TD, it is critical to identify and manage PTD as early as possible, which can make developers aware of the existence of PTD, and then decide when and how to resolve the PTD to avoid the occurrence of TD. Nowadays, many research and industrial tools have been provided (such as static code analysis tools) to detect TD, but these tools lack consensus~\cite{lefever2021lack} and a large amount of PTD cannot be detected by existing tools 
\cite{kruchten2012technical}.

 
Considering this, \textbf{we focused on} the concept of PTD and conducted an exploratory study to understand PTD and its distributions, classification, and resolution, as identified in code reviews. We first collected code review comments from two widely known projects: the Nova project of OpenStack and the Qt Base project of Qt. We then manually labelled and analyzed each review comment to study how much PTD is identified in code reviews, what types of PTD are identified, what actions are taken by developers to resolve the PTD, and how long it takes for developers to pay back the TD. We analyzed 163 PTD-related review comments acquired from a total of 2,030 review comments from the two projects. 

The remaining sections of this paper are structured as follows: Section \ref{related-work} surveys the related work on TD and code reviews. Section \ref{Research Design} describes the research design of this study. The results of our study are presented in Section \ref{Results}, followed by a discussion of the implications of our results in Section \ref{Implications}. Section \ref{Threats to validity} clarifies the threats to the validity. Finally, we conclude this work in Section \ref{Conclusions}.

\section{Related Work} \label{related-work}
\subsection{Technical Debt}
According to Li \textit{et al}. \cite{li2015systematic}, TD occurs when technical shortcuts are taken by developers to gain short-term profits that are harmful to the system in the long term. In recent years, the research community has extensively worked on different aspects of TD. Zazworka \textit{et al}. compared the results between manual identification of TD by researchers and automatic detection of TD by tools \cite{zazworka2013case}. Their result shows that the tools employed are especially useful for identifying defect debt but cannot help in identifying many other types of debt, so involving human effort in the identification process is necessary. De Toledo \textit{et al}. conducted a case study on a project from a large company to investigate architecture TD in microservices architecture \cite{de2019architectural}. They combined the existing documentation and interviews to detect problems, solutions, and risks, providing a list of architectural issues that generate TD, which is useful for practitioners that want to avoid or repay TD. 

Moreover, many studies focus on identifying TD from various types of sources. For instance, Potdar \textit{et al.} used code comments in four large Open Source Software (OSS) projects to identify SATD. They found that developers with higher experience tend to introduce most of the SATD \cite{potdar2014exploratory}. Silva \textit{et al.} focused on identifying TD in pull requests, and their results reveal that 30.3\% of the pull request rejected are due to TD, and the most frequently TD encountered is design TD \cite{silva2016does}. Li \textit{et al.} explored SATD in issue trackers, and found that the majority of TD is paid off \cite{9226330}. 

Inspired by the body of work discussed above, we were motivated to investigate PTD, a source of TD \cite{6608681}. The difference between our work and existing work is that we identified and analyzed PTD in the context of code review, which involves reliable human identification process (i.e., reviewers could consider taking full consideration of the contextual information and thus be better positioned to detect PTD in the code). Recently, an empirical study was conducted by Kashiwa \textit{et al}. to examine the relationship between SATD and code review process, which is the most related work to us \cite{KASHIWA2022106855}. However, in that work, the authors mined SATD only from the code comments that introduce SATD during the code review process. Moreover, the paper did not investigate the repayment or resolution of TD in code review. Specifically, we considered identifying PTD from the review comments of code reviewers and developers to understand PTD and its distribution, classification, and resolution in code reviews.


\subsection{Code Review}
Code review is a typical process in software development. Many studies have explored modern code review processes in practice. For instance, McIntosh \textit{et al.} studied the relationship between code review and post-release defects \cite{mcintosh2016empirical}. Their results empirically confirm the intuition that poorly-reviewed code has a negative impact on software quality. Baysal \textit{et al.} investigated the influence of non-technical factors (such as patch size, organizational and personal dimensions) on code review \cite{6671287}. Their results show that organizational and personal factors influence review timeliness, as well as the likelihood of a patch being accepted.  

There are also studies focusing on investigating a variety of artifacts in code reviews. For instance, Zanaty \textit{et al.} set out to study the frequency and nature of design discussions in code reviews \cite{designDiscussion}. They collected and analyzed the review comments of the OpenStack's Nova and Neutron projects, and their results indicate that design concerns are not commonly discussed in code reviews and most design related comments are constructive. Nanthaamornphong \textit{et al.} examined comments from code reviewers that identified code smells in two open source projects (i.e., OpenStack and WikiMedia) \cite{nanthaamornphong2016empirical}. They preliminary results show that code reviewers comment on only a small number of code smells. In order to further investigate code smells in code reviews, Han \textit{et al}. conducted an empirical study using the same projects as used by Zanaty \textit{et al.} \cite{designDiscussion} (Nova and Neutron) to study how code smells are identified in code reviews, and the actions taken against the identified smells \cite{hanCodeSmell, han2022code}. Their result shows that reviewers usually provided constructive feedback, including fixing (refactoring) recommendations to help developers remove smells. Mello \textit{et al.} focused on exploring the influence of human factors for identifying code smells in code reviews \cite{8170086}. Their results reveal that human factors play a key role in the precise identification of code smells, and reviewers with professional background could reach a high precision of smell identification. 

The above work motivates us to identify and analyze PTD in the context of code reviews. Their selection of communities and projects lay the foundation for this study. 

\section{Research Design} \label{Research Design}

\subsection{Research Questions}
The goal of this work, formulated through a Goal Question Metric approach \cite{caldiera1994goal}, is to investigate the existence of PTD in code reviews \textbf{for the purpose of} exploration \textbf{with respect to} PTD and its distribution, classification, and resolution \textbf{from the point of} developers and code reviewers \textbf{in the context of} OSS projects. To achieve this goal, we formulated the following five research questions (RQs), as follows:

\noindent\textbf{RQ1. What is the proportion of review comments that indicate PTD in code reviews?}\\
\textbf{Rationale:} As an exploratory study on PTD in code reviews, this RQ aims at providing a basic view of the distribution of PTD, which can help improve developers' awareness of PTD. 

\noindent\textbf{RQ2. What are the types of PTD identified in code reviews?}
\textbf{Rationale:} Different types of PTD may have different impact on development, and they may also have different priorities to be resolved. Therefore, this RQ intends to provide an understanding regarding the classification of PTD in code reviews. Moreover, by providing the distribution of different types of PTD in code reviews, the results of this question can also help developers be aware of the most common problems that they are confronted with. 

\noindent\textbf{RQ3. How much of the identified PTD has been resolved and how much of the identified PTD has evolved into TD?}\\
\textbf{Rationale:} When PTD is identified in the code reviews, developers could decide whether resolve it in the later patchsets of the code change or temporarily ignore it. In the latter case, the PTD will evolve into TD because it is not resolved when code change is merged into the main code repository, which can have a negative impact on the software system. To help stakeholders better manage PTD, it is necessary to first quantify how much of the identified PTD has been resolved or evolved into TD.

\noindent\textbf{RQ4. How long does it take for developers to resolve identified PTD in code reviews?}\\
\textbf{Rationale:} For the PTD that has been resolved, we would like to know how long it takes for developers to resolve the identified PTD. Answering this RQ can help better prepare developers for fixing similar issues in the future. In addition, since the severity and resolution difficulty of various types of PTD can be different, we further investigate the difference in time taken when resolving different types of PTD.

\noindent\textbf{RQ5. What practices are used by developers to resolve PTD that has been identified in code reviews?}\\
\textbf{Rationale:} By answering this RQ, we aim to investigate the practices employed by developers to resolve PTD. Such information could help stakeholders (e.g., developers) better manage and resolve PTD in the development process.

\subsection{Data Collection}
This study analyses PTD in code reviews collected from two large projects of the OpenStack\footnote{\url{https://www.openstack.org/}} and Qt\footnote{\url{https://www.qt.io/}} communities. OpenStack is a set of software tools for building and managing cloud computing platforms, and is supported by many large companies. Qt is a cross-platform application and UI framework developed by the Digia corporation, but welcomes contributions from the community at large. Since these two communities have made a big investment in code reviews for several years \cite{HiraoCode} and are widely used in many studies related to code reviews \cite{6624003, wang2021understanding}, we deemed them to be appropriate and representative for our analysis. The OpenStack and Qt communities are composed of several projects, and we selected one of the most active projects from each community (based on the highest number of closed code changes), i.e., Nova\footnote{\url{https://github.com/openstack/nova}} from OpenStack and Qt Base\footnote{\url{https://github.com/qt/qtbase}} from Qt.

OpenStack and Qt adopt Gerrit\footnote{\url{https://www.gerritcodereview.com/}}, a web-based code review platform built on top of Git, to support their code review process. By using the RESTful API provided by Gerrit, we collected all the code changes of the two selected projects that were last updated in 2020. We then extracted all available review comments of these code changes and stored the data in a local file for further analysis. In total, we collected a dataset of 1,840 code changes and 13,982 review comments from Nova, and 8,954 code changes and 36,924 review comments from Qt Base.

\subsection{Data Labelling and Extraction}\label{Data Labelling and Extraction}

Manually analyzing all the review comments from Nova (13,982) and Qt Base (36,924) is a time-consuming task. Considering this, we decided to randomly select a representative sample of review comments for each project based on 3\% margin of error and 95\% confidence level~\cite{israel1992dss}. This leads to the selection of 992 and 1,038 review comments from Nova and Qt Base, respectively. 

In the \textbf{data labelling} step, we manually read through all review comments to label whether they are related to PTD. For the PTD-related review comments, we classified and labelled their types using the TD taxonomies provided by Maldonado \textit{et al.} \cite{MaldonadoDetect}, Liu \textit{et al.} \cite{tdindl}, and Li \textit{et al.} \cite{li2015systematic}. The reason we chose the description of TD types as the basis for classifying PTD is that PTD will evolve into the corresponding type of TD when it is injected into the software systems without being resolved. 

To mitigate any selection bias, we conducted a pilot labelling of the PTD-related comments with all the 992 code review comments randomly selected from Nova by the first and third authors, independently. The disagreements on the labelling results were discussed and resolved with the second author to get a consistent understanding of the criteria of data labelling, i.e., the review comment should be clearly related to PTD and meet the definition of one of the TD types, which are illustrated in Section \ref{Results}. After the first and third authors achieved a consensus on the labelling of all the review comments of Nova, they then further labelled the 1,038 code review comments randomly selected from Qt Base, independently. Any disagreements on the labelling results were still discussed and resolved with the second author. We measured the inter-rater reliability and calculated the Cohen’s Kappa coefficient \cite{cohen1960coefficient} as a way to verify the consistency on the labelled review comments of Qt Base between the first and third authors. The Cohen's Kappa coefficient is 0.81, indicating that the two coders reached a decent agreement on labelling PTD-related comments.

\begin{figure*}
    \centering
    \includegraphics[width=1\textwidth]{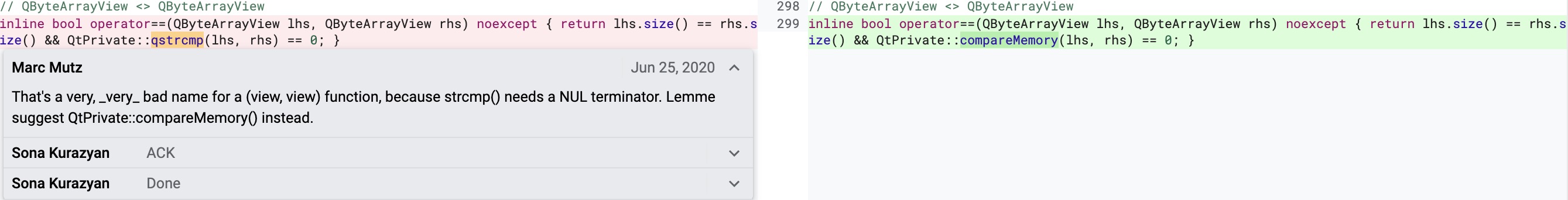}
    \caption{An example of a \textit{renaming bad name} operation to resolve PTD after review (the change is highlighted)}
    \label{code change}
\end{figure*}

In the \textbf{data extraction} step, for answering RQ3, RQ4, and RQ5, we analyzed the contextual information of each identified PTD-related review comment, including the code review discussions and associated source code to determine whether the identified PTD was resolved. Specially, we regarded a PTD as resolved when the resolution situation fits into the following three categories:

\begin{enumerate}
\item[1)] Changes were made in the code by the developer(s) to resolve the PTD before the code change is merged (see Figure \ref{code change}).

\item[2)] Developer(s) clearly mentioned that the PTD has been resolved in another follow-up code change. 

\item[3)] The code changes that introduced the PTD were abandoned so that the PTD would not have any effect on the main code repository. In other words, the PTD no longer exists.
\end{enumerate}

\looseness-1  For each resolved PTD, we extracted and recorded its resolution information. Specifically, for the existence time of PTD, we calculated the time interval between the identification time (the timestamp of when the corresponding review comment was written) and the resolution time (the timestamp of when developers committed patchset to resolve the PTD or when developers abandoned the code change that introduced the PTD). The resolution practices were analyzed based on the extracted information using Constant Comparison \cite{glaser1965constant}. This data extraction process was conducted by the first author and the results were verified by two other co-authors. Conflicts were discussed and addressed by the three authors. Table \ref{data items} presents the data items to be extracted and their corresponding RQs in this study, which have been provided online \cite{replication-package}. 



\begin{table}[h]
\small
\caption{Data Items to be Extracted from Review Comments}
\resizebox{\columnwidth}{!}{
\begin{tabular}{|l|l|l|}
\hline
\textbf{Data Item}  & \textbf{Description}                                                                                               & RQ  \\ \hline
PTD-Related         & \begin{tabular}[c]{@{}l@{}}Whether the review comment is related \\ to PTD (i.e., Yes or No).\end{tabular}         & RQ1 \\ \hline
PTD                 & The PTD text extracted from the message.                                                                           & RQ1 \\ \hline
Who Identifies PTD &
  \begin{tabular}[c]{@{}l@{}}The role of person who identifies PTD in  \\ the review comment (i.e., Reviewer, \\ Developer, or Developer after discussion).\end{tabular} &
  RQ1 \\ \hline
PTD-Type            & The type of the identified PTD.                                                                                    & RQ2 \\ \hline
Resolved            & \begin{tabular}[c]{@{}l@{}}Whether the identified PTD is resolved \\ (i.e., Yes or No).\end{tabular}               & RQ3 \\ \hline
Resolution Evidence & \begin{tabular}[c]{@{}l@{}}Any information about where the \\ identified PTD is resolved in the code.\end{tabular} & RQ3 \\ \hline
Identification Time & \begin{tabular}[c]{@{}l@{}}The time when the review comment that \\ identifies PTD is added.\end{tabular}          & RQ4 \\ \hline
Resolution Time     & The time when identified PTD is resolved.                                                                          & RQ4 \\ \hline
Existence Time &
  \begin{tabular}[c]{@{}l@{}}The time interval between identification \\ time and resolution time of the identified PTD.\end{tabular} &
  RQ4 \\ \hline
Resolution Practice & \begin{tabular}[c]{@{}l@{}}The practice that developers take to resolve \\ the identified PTD.\end{tabular}        & RQ5 \\ \hline
\end{tabular}
}
\label{data items}
\end{table}

\section{Results and Analysis} \label{Results}

\textbf{RQ1: What is the proportion of review comments that indicate PTD in code reviews?}\\
Table \ref{Overview} presents an overview of review comments per project. In general, we identified a total of 163 review comments that indicate PTD. Compared with the number of all the review comments we analyzed, we can find that PTD is not so prevalent in code reviews, only accounted for 8.0\% on average. Concretely, there are 90 PTD-related review comments that are identified in code reviews for Nova (out of 992, 9.1\%). The number of PTD-related review comments identified from Qt Base is less than that of Nova, and the proportion is a bit lower (73 out of 1,038, 7.0\%). 

We further investigated how much of PTD is identified by code reviewers and developers. We categorized the role of the person who identified PTD in code reviews into three situations, i.e., Reviewer, Developer, or Developer after discussion. We found that developers sometimes take the initiative to write down the review comments and record the issues that contain PTD in code reviews. However, it is not so prevalent with only 4 instances (out of 163, 2.5\%). In contrast, most of PTD was identified by reviewers (143 out of 163, 87.7\%). Moreover, PTD can also be identified in the response comments of developers. In this situation, developers can be aware of the existence of PTD through the discussion with reviewers (16 out of 163, 9.8\%). As shown in the example below, the developer forgot to remove the redundant code after debugging and was aware of the code issue after it was pointed out by the reviewer, which indicates a code PTD. In a word, the identification of the overwhelming majority of PTD (97.5\%) owes to the help of reviewers in the code review process. This finding indicates that developers may not realize that they introduce the PTD during the development process. It also shows the important role that reviewers play in the code review process for identifying the PTD.

\FrameSep=8pt\FrameRule=1pt\begin{framed}
\noindent\textbf{Link:} \url{http://alturl.com/o24nq} \\
\textbf{Project:} Nova \\
\textbf{Reviewer:} ``I don't get this... But then why we register them for this test.'' \\
\textbf{Developer:}``Sorry, this was debug ... I added this to make it run consistently in isolation while I figured out what was going on. I meant to remove this and forgot.''
\end{framed}

\FrameSep=3pt\FrameRule=0.5pt\begin{framed}
\noindent  \textbf{RQ1 Summary:} PTD can be identified in code reviews but is not prevalent. Specifically, only 8.0\% of review comments contain PTD on average. Moreover, reviewers play a significant role for identifying PTD in the code review process.
\end{framed}
 
\begin{table}[h]
\small
\caption{Overview of Review Comments per Project}
\resizebox{0.47\textwidth}{!}{%
\begin{tabular}{|l|l|l|l|}
\hline
\textbf{Project}        & \textbf{Review Comments} & \textbf{PTD-related Review Comments} & \textbf{\%}   \\ \hline
Nova            & 992             & 90                         & 9.1 \\ \hline
Qt Base         & 1038            & 73                         & 7.0 \\ \hline
\textbf{Total}  & 2030            & 163                        & 8.0 \\ \hline
\end{tabular}%
}
\label{Overview}
\end{table}


\textbf{RQ2: What are the types of PTD identified in code reviews?}\\
As mentioned in Section \ref{Data Labelling and Extraction}, we adopted the classification of TD types provided in Maldonado \textit{et al.} \cite{MaldonadoDetect}, Liu \textit{et al.} \cite{tdindl}, and Li \textit{et al.} \cite{li2015systematic} as the basis for classifying PTD in this work. Moreover, the aforementioned three studies also provide detailed \textit{causes} of TD, which are helpful for understanding and classifying PTD.

In this work, we identified six types of PTD in code reviews, and we provide examples of review comments for each PTD type below.

\textbf{1. Design PTD} refers to the technical shortcuts that are taken in detailed design, which lead to sub-optimal design of the code. Various reasons could lead to design PTD, e.g., lack of abstraction, poor implementation, workarounds, or temporary solutions. 

We found that \textit{poor implementation of code} \cite{MaldonadoDetect, tdindl} is the most important reason that leads to design PTD in code reviews. In this case, the current implementation may satisfy the functionality, but developers are unaware that the design of the code is sub-optimal. 

\FrameSep=8pt\FrameRule=1pt\begin{framed}\noindent
\textbf{Link:} \url{http://alturl.com/6srnp} \\
\textbf{Project:} Nova \\
\textbf{Reviewer:} ``Don't do this at query time. There are not likely to be many of these and apparently these queries are \texttt{\_slow\_}. Instead, do it in the Python code below.''
\end{framed}

\FrameSep=8pt\FrameRule=1pt\begin{framed}\noindent
\textbf{Link:} \url{http://alturl.com/s63p7} \\
\textbf{Project:} Qt Base \\
\textbf{Reviewer:} ``This leads to very poor code expansion. is\_shared should either be LSB or MSB. The advantage of being MSB (last in declaration order for little-endian) is that it becomes the sign bit.''
\end{framed}

The reviewers were negative about current implementation and recommended specific implementation to improve the design of the code. Due to the lack of experience or context knowledge of the projects, developers might write not-quite-right code without considering whether the implementation is optimal. Such poor implementation can lead to various quality issues in the software systems (e.g., low performance and extensibility in our examples).

\textit{Code duplication} \cite{tdindl} is another reason that can result in design PTD in code reviews. In order to reduce the workload in development, developers aim at reusing code as much as possible. To achieve this goal, for example, methods with single responsibility can be extracted and reused rather than duplicating the same code in multiple places. In the below example, the reviewer thought that the developer should write a helper function in the code to meet the DRY (Don't Repeat Yourself) principle. Code duplication could increase the maintenance burden since developers might need to change the code in multiple places to reflect on a single change elsewhere in the code. This in turn can lead to bugs being introduced if part of code is left unchanged. 

\FrameSep=8pt\FrameRule=1pt\begin{framed}\noindent
\textbf{Link:} \url{http://alturl.com/wgyyu} \\
\textbf{Project:} Nova \\
\textbf{Reviewer:} ``Shouldn't we have kept the helper function (and modify and rename it) for doing this part to keep this DRYer?''
\end{framed}

\textbf{2. Defect PTD} indicates that the code behaves in an unexpected way since there are defects/bugs in the code. 
As shown in the example below, the developer mistakenly thought that the code works without any issues. However, the reviewer pointed out that the wrong code is potentially flaky (behaves non-deterministically), indicating that there is a defect in the code. 

\FrameSep=8pt\FrameRule=1pt\begin{framed}\noindent
\textbf{Link:} \url{http://alturl.com/a2pap} \\
\textbf{Project:} Qt Base \\
\textbf{Reviewer:} ``We shouldn't call qui methods from non-gui threads I guess.'' \\
\textbf{Developer:} ``Not sure what's different with QStatusBar, but it works without any issues ...'' \\
\textbf{Reviewer:} ``Well, you are apparently calling showMessage from the different thread than statusBar lives in. QStatusBar is not reentrant, neither showMessage. Works just by coincidence, but it's wrong.''
\end{framed}

\textbf{3. Documentation PTD} indicates the \textit{lack of code comments} \cite{li2015systematic} or \textit{inadequate documentation} \cite{tdindl, li2015systematic} that could explain the corresponding part of the system. 
We found that reviewers usually expect developers to provide clear descriptions of the code they write by providing code comments. Nevertheless, developers sometimes forget to write code comments, which leads to the documentation PTD. For example:

\FrameSep=8pt\FrameRule=1pt\begin{framed}\noindent
\textbf{Link:} \url{http://alturl.com/p75ju} \\
\textbf{Project:} Nova \\
\textbf{Reviewer:} ``supernit - I know it's pretty straight forward but a quick comment outlining what the test is doing would be useful.''
\end{framed}


Software documentation is a critical activity in software engineering \cite{isd}. Proper documentation can not only make it easier for other developers to reuse the code, but also decrease the costs and difficulties of maintenance. Developers should attach great importance to documentation and write high quality documentation for the code they write.

We also found that \textit{outdated documentation} \cite{li2015systematic} will lead to documentation PTD. In the below example, the reviewer pointed out that the function documentation should be updated since the function has been changed, which indicates that the old code comment was outdated. Outdated documentation can mislead the readers so that they may fail to grasp the intent of the code. If other developers are misled by outdated documentation when they plan to reuse the code, it may lead to code issues with unexpected costs. 

\FrameSep=8pt\FrameRule=1pt\begin{framed}\noindent
\textbf{Link:} \url{http://alturl.com/any97} \\
\textbf{Project:} Nova \\
\textbf{Reviewer:} ``This function now returns one single \texttt{ResourceProvider} hence the function doc needs to be updated''
\end{framed}

\textbf{4. Requirement PTD} indicates the incompleteness of method, class, or program. 

The reviewers would point out the issues if they think the code is incomplete. Due to time pressure or other constraints, developers sometimes did not immediately follow reviewers' recommendations but chose to leave a \texttt{TODO} comment to remind themselves to do it later. Therefore, this can be a requirement PTD until the \texttt{TODO} comment is completed. For example:

\FrameSep=8pt\FrameRule=1pt\begin{framed}\noindent
\textbf{Link:} \url{http://alturl.com/8czn4} \\
\textbf{Project:} Nova \\
\textbf{Reviewer:} ``I would raise an exception if a key already defined to help avoid duplicated validators.'' \\
\textbf{Developer:} ``Yup, added a \texttt{TODO} to do this. This whole thing is a bit janky at the moment and needs to be shuffled around.''
\end{framed}

Moreover, we observed that new requirements can be identified in the discussions between reviewers and developers. In the below example, we can see that a new functional requirement is mentioned by the developer (i.e., ``\textit{a future improvement on QFutureInterface to make it movable}''). In this context, the new requirement is clearly known by the developer, and hence should be scheduled in the future development plan. Therefore, this can also be a requirement PTD until the new requirement is fulfilled.

\FrameSep=8pt\FrameRule=1pt\begin{framed}\noindent
\textbf{Link:} \url{http://alturl.com/eu4zw} \\
\textbf{Project:} Qt Base \\
\textbf{Reviewer:} ``\texttt{d(std::exchange(other.d, {}))}?'' \\
\textbf{Developer:} ``Oh, good suggestion. There's no move semantics in \texttt{d} at the moment (unsure how \texttt{std::exchange} requires MoveConstructible to be true), but that's a topic of future improvement: making \texttt{QFutureInterface} movable and simplifying both this part and swap.''
\end{framed}

\textbf{5. Test PTD} indicates the need for implementation or improvement of the current tests. We found that \textit{lack of tests} \cite{tdindl, li2015systematic} is the main reason that leads to test PTD. 

As shown in the below examples, the review comments related to test PTD explicitly convey the meaning that tests are insufficient and the developers should add new tests or improve existing tests. Lack of tests makes it hard for developers to find bugs and defects buried in the code, and then increases the number of defect PTD.

\FrameSep=8pt\FrameRule=1pt\begin{framed}\noindent
\textbf{Link:} \url{http://alturl.com/obj6s} \\
\textbf{Project:} Nova \\
\textbf{Reviewer:} ``This seems not covered by test case.''
\end{framed}

\FrameSep=8pt\FrameRule=1pt\begin{framed}\noindent
\textbf{Link:} \url{http://alturl.com/fckqu} \\
\textbf{Project:} Qt Base \\
\textbf{Reviewer:} ``there must be tests for terminate \& kill - these seem more relevant to this matter, imo. feel free to fix the test in a separate commit, obviously''
\end{framed}

\textbf{6. Code PTD} refers to the poorly written code that violates best coding practices or coding conventions.

We found that \textit{violating code conventions} \cite{li2015systematic} is one of the reasons that can lead to code PTD identified in code reviews. Taking bad naming as an example, following a naming convention plays an important role in making code readable. In the below example, the code reviewer identified an issue of naming variables with a single character (i.e., \texttt{f} and \texttt{l}). A meaningful naming should tell you what the variable or function stands for. Otherwise, this might confuse those reading the code. That is why the reviewer recommended to rename \texttt{f} and \texttt{l} to \texttt{firstIter} and \texttt{lastIter}, respectively.  

\FrameSep=8pt\FrameRule=1pt\begin{framed}\noindent
\textbf{Link:} \url{http://alturl.com/9zog3} \\
\textbf{Project:} Qt Base \\
\textbf{Reviewer:} ``Please try harder; single-char identifiers are unreadable, and \texttt{l} is particularly bad for reading. firstIter/lastIter will work sufficiently well.''
\end{framed}

We also found that \textit{low-quality code} \cite{li2015systematic} identified in code reviews can lead to code PTD. There are various reasons that decrease the quality of the code, e.g., bad copy/paste activities, redundant code, using magic value, and so on. For instance, developers sometimes copy and paste code snippets without thinking about whether the code is appropriate to reuse in their context, which decreases the quality of the code in general, as in the following example:

\FrameSep=8pt\FrameRule=1pt\begin{framed}\noindent
\textbf{Link:} \url{http://alturl.com/y6zmn} \\
\textbf{Project:} Nova \\
\textbf{Reviewer:} ``
I suspect it was more a case of copying prior art from `\texttt{nova.tests.functional.regressions}'.''
\end{framed}

Moreover, as shown in the example below, we can see that developers sometimes use magic value in their code, which decreases the quality of the code because using magic value can degrade the readability of the code and make it harder to maintain.

\FrameSep=8pt\FrameRule=1pt\begin{framed}\noindent
\textbf{Link:} \url{http://alturl.com/vwdcw} \\
\textbf{Project:} Qt Base \\
\textbf{Reviewer:} ``Please duplicate the code from above with the comment to avoid the 1 magic value: ...''
\end{framed}

Table \ref{dis_type} presents the distribution of different types of PTD in code reviews ordered by their numbers. We can find that code PTD is by far the most frequently identified PTD, with exactly 55 instances (33.7\%). The second most frequent type is documentation PTD, which accounts for 27.6\%, nearly the same as code PTD. There are 21 review comments (12.9\%) that identified test PTD, followed by design PTD making up 12.3\%. Requirement PTD and defect PTD are the types with the lowest proportions (less than 10\%).

\begin{table}[h]
\small
\caption{Distribution of Types of PTD in Code Reviews}
\begin{tabular}{@{}|l|l|r|@{}}
\hline
\textbf{Type}            & \textbf{Number} & \textbf{Percentage}      \\ \hline
Code PTD          & 55     & 33.7\%  \\ \hline
Documentation PTD & 45     & 27.6\%  \\ \hline
Test PTD          & 21     & 12.9\%  \\ \hline
Design PTD        & 20     & 12.3\%  \\ \hline
Requirement PTD   & 12     & 7.4\%   \\ \hline
Defect PTD        & 10     & 6.1\%   \\ \hline
\end{tabular}
\label{dis_type}
\end{table}

\FrameSep=3pt\FrameRule=0.5pt\begin{framed}
\noindent \textbf{RQ2 Summary:} We have identified six types of PTD in code reviews, i.e., design, defect, documentation, requirement, test, and code PTD. The majority of the PTD identified in code reviews is code PTD (33.7\%) and documentation PTD (27.6\%).
\end{framed}


\textbf{RQ3: How much of the identified PTD has been resolved and how much of the identified PTD has evolved into TD?}\\
Table \ref{payment} presents the resolution situation of the PTD identified in code reviews. Of the 163 PTD, the majority (132, 81.0\%) has been resolved by the developers after the review. In detail, the proportions of the TD that was resolved in Nova and Qt Base are almost the same, which are 81.1\% and 80.8\%, respectively. This finding indicates that developers tend to be resolve the PTD under the suggestions of reviewers to help increase the possibility of code changes passing the code review. However, 19.0\% of PTD remains unresolved.  Therefore, once the code change is merged, the unresolved PTD would be injected into the software system in the form of TD (i.e., evolves into a real TD), which might have serious negative impact on the main code repository.

\begin{table}[h]
\small
\caption{Resolution Situation of Each Project }
\begin{tabular}{|l|l|l|l|l|}
\hline
Project        & Resolved & \% Resolved & Remaining & \% Remaining \\ \hline
Nova           & 73       & 81.1\%      & 17        & 18.9\%       \\ \hline
Qt Base        & 59       & 80.8\%      & 14        & 19.2\%       \\ \hline
\textbf{Total} & 132      & 81.0\%      & 31        & 19.0\%       \\ \hline
\end{tabular}
\label{payment}
\end{table}

Moreover, we also investigated the resolution situation of each PTD type as shown in Table \ref{payment_type}. From the table, we can see that test PTD is the debt with the highest percentage of being paid by developers, which corresponds to 90.5\%. We can also learn that the resolution proportion of code PTD is also very high (close to test PTD) even if the number of code PTD is the highest. This finding indicates that developers show their great emphasis on code TD, and they resolve code PTD with an active attitude. The resolution proportion of documentation PTD is also quite high, which is 86.7\%. This may be attributed to the explicit targets of how to resolve documentation PTD, e.g., developers could supplement or update the corresponding code comments after reviewers point out what comments are missing or which documentation is outdated. However, we can notice that only 16.7\% of requirement PTD has been resolved by developers, which is much lower than other types of PTD. Potential reasons could be that: (1) a requirement may rely on another requirement, and thus it cannot be implemented in time. As shown in the response of a developer ``\textit{would you be ok with me just adding a TODO in the code to extend the virt driver rebuild function if support is added for cyborg to a driver that uses it?}'', the developer could only extend the rebuild function after adding support for cyborg; (2) many requirements are related to the further improvement on methods, classes, or program. Therefore, compared with other issues identified in code reviews, such requirements may be put in lower priorities by developers and be schedules and fulfilled in the future.

\begin{table}[h]
\small
\caption{Resolution Situation of Each PTD Type}
\begin{tabular}{@{}|l|l|l|r|@{}}
\hline
\textbf{Type}            & \textbf{Num} & \textbf{Resolved by Developers} & \textbf{\%} \\ \hline
Code PTD          & 55     & 49                 & 89.1\%        \\ \hline
Documentation PTD & 45     & 39                 & 86.7\%        \\ \hline
Test PTD          & 21     & 19                 & 90.5\%        \\ \hline
Design PTD        & 20     & 16                 & 80.0\%        \\ \hline
Requirement PTD   & 12     & 2                  & 16.7\%        \\ \hline
Defect PTD        & 10     & 7                  & 70.0\%        \\ \hline
\end{tabular}
\label{payment_type}
\end{table}

\FrameSep=3pt\FrameRule=0.5pt\begin{framed}
\noindent \textbf{RQ3 Summary:} On average, about 81.0\% of the PTD identified in code reviews was resolved by developers while remaining 19.0\% of the PTD evolved into the TD. Moreover, most of test PTD, code PTD, and documentation PTD have been resolved by developers. 
\end{framed}
%

\begin{figure}[h]
    \centering
    \includegraphics[width=0.45\textwidth]{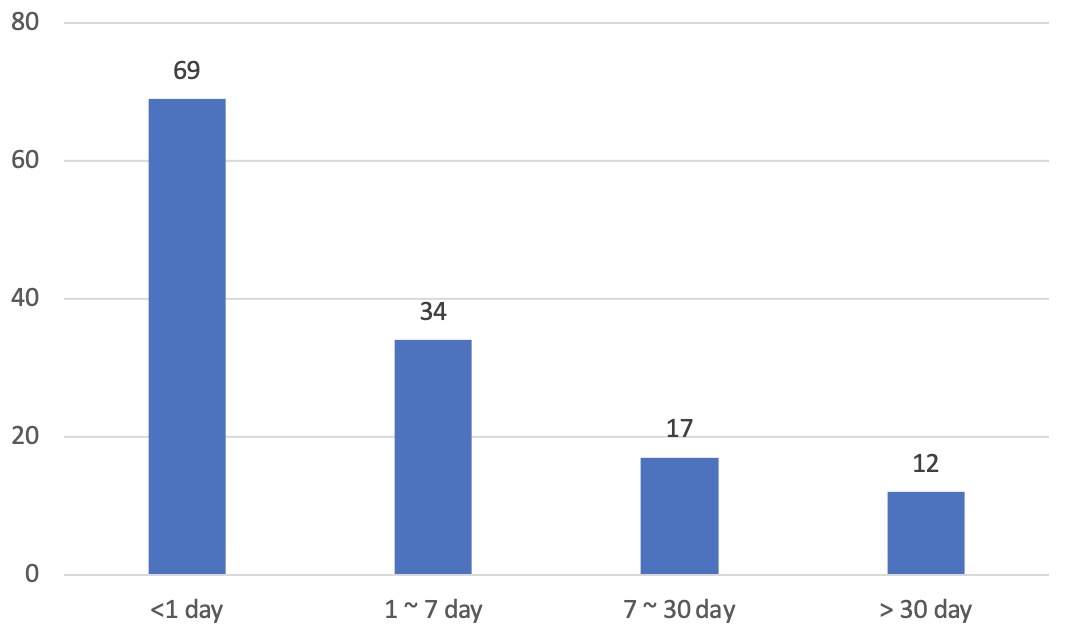}
    \caption{Distribution of Time for Resolution}
    \label{pie chart}
\end{figure}

\textbf{RQ4: How long does it take for developers to resolve identified PTD in code reviews?}\\
According to the results of RQ3, a total of 132 PTD identified in code reviews was resolved by developers. To show the distribution of time for resolution, we plotted the results in a histogram (as shown in Figure \ref{pie chart}). From the figure, we can see that more than half of PTD has been resolved in less than one day, which corresponds to 69 (52.3\%) of 132 PTD-related review comments. In 34 (25.8\%) PTD-related review comments, developers chose to resolve the PTD in more than one day but less than one week after review. That is to say, developers can resolve the majority of the PTD in less than a week (103 out of 132, 78.0\%). 17 (12.9\%) of the PTD takes developers 2-4 weeks to resolve, and only few PTD (9.1\%) costs developers more than a month to fix. 

Moreover, we also investigated the resolution time of each PTD type. As shown in the Table \ref{time of PTD type}, we listed the minimum and median time for resolving different types of PTD. We find that the minimum time taken for resolving all types of PTD except requirement PTD can be very short - only near one hour. Although the number of the resolved requirement PTD is limited in our cases, the longest minimum resolution time of requirement PTD (289 hours) still can reflect the hardness when developers were confronted with requirement issues. According to the median results, we can find that the documentation PTD (median = 18 hours) was resolved faster than other types of PTD, which suggests that the documentation PTD would be easier to be addressed by developers. We believe this is due to the nature of documentation PTD. Usually the targets of how to resolve documentation PTD are rather explicit as described in the results of RQ3, and there is no need to modify the code when resolving the documentation PTD. Code PTD (median = 22 hours) and Design PTD (median = 27 hours) are also quite fast to resolve, which costs around a day. The resolution time of Defect PTD (median = 48 hours) and Test PTD (median = 116 hours) is much longer than the aforementioned three types of PTD (i.e., documentation, code, and design PTD). Same as the minimum result of requirement PTD, the median result of requirement PTD (325.5 hours) also reveals the fact that developers need to take more time to consider about how to further implement and complete the incomplete method, class, or program.

\begin{table}[h]
\caption{Time Taken for Resolving Different Types of PTD}
\begin{tabular}{|l|l|l|}
\hline
\textbf{Type}     & Minimum (hour) & Median (hour) \\ \hline
Code PTD          & 1h   & 22h     \\ \hline
Documentation PTD & 1h   & 18h    \\ \hline
Test PTD          & 1h   & 116h    \\ \hline
Design PTD        & 1h   & 27h     \\ \hline
Requirement PTD   & 289h & 325.5h  \\ \hline
Defect PTD        & 1h   & 48h     \\ \hline
\end{tabular}
\label{time of PTD type}
\end{table}


\FrameSep=3pt\FrameRule=0.5pt\begin{framed}
\noindent \textbf{RQ4 Summary:} On average, about 78.0\% of the resolved PTD was resolved by developers within a week. Moreover, documentation PTD takes less time to resolve (18 hours on median) while requirement PTD costs the longest time to resolve (325.5 hours on median).
\end{framed}
%

\textbf{RQ5: What practices are used by developers to resolve PTD that has been identified in code reviews?} \\
For answering this question, we analyzed the code review discussions and associated source code of each PTD, and identified the practices used by developers to resolve the PTD identified in code reviews. Five resolution practices are then identified, which are presented below:

\begin{enumerate}
\item[1)] \textbf{Code refactoring} is the process of restructure the code with the aim of improving the design, structure, or the non-functional properties (e.g., readability, maintainability, or complexity) while preserving its functionality.
\item[2)] \textbf{Documentation improvement} is the process of improving the quality of documentation, including supplementing code comments, restructuring the statements of documentation, or removing the outdated documentation. 
\item[3)] \textbf{Testing improvement} is the process of adding new tests or improving the existing tests.
\item[4)] \textbf{Bug fixing} is the process of correcting the known defects/bugs in the software system.
\item[5)] \textbf{Code change abandonment} refers to the cases where developers may abandon the code changes. In such cases, the code that contains PTD will not be merged in the main code repository so that the PTD will not have an effect on the project. Considering this, we regarded it as a special resolution practice. However, the intentions of developers to abandon the code changes may not be necessarily due to PTD. Therefore, we treated it as a resolution practice only if developers did not use one of the other four resolution practices and the code change was abandoned.

\end{enumerate}

\begin{figure}[h]
    \centering
    \includegraphics[width=0.45\textwidth]{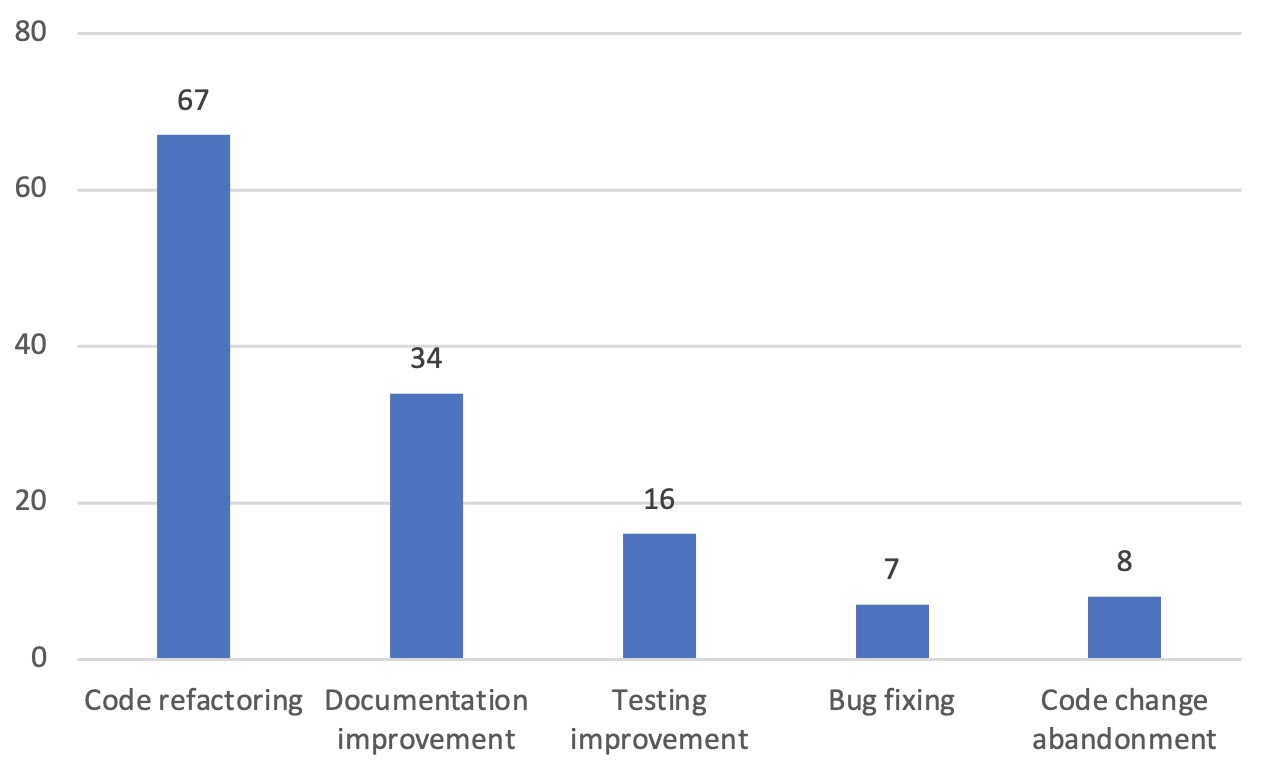}
    \caption{Distribution of Resolution Practices}
    \label{payment_practice}
\end{figure}

As demonstrated in Figure \ref{payment_practice}, we can see that the most frequent resolution practice used by developers is \textbf{Code refactoring} (67 out of 132, 50.9\%), which is almost two times greater than the second most used resolution practice in our cases. Some of the examples replied by developers to the review comments are: ``\textit{This code would need a refactoring to become clearer IMHO.\footnote{\url{http://alturl.com/k4qhi}}}'' and ``\textit{Refactored in PS8.\footnote{\url{http://alturl.com/8ptv7}}}''. Code refactoring is a base practice that can be used for a variety of purposes, such as improving the design of an existing code base \cite{fowler2018refactoring}, improve the internal quality of the code \cite{li2015systematic}, and so on. Therefore, it plays an important role in resolving code PTD and design PTD identified in code reviews.

With the high proportion (27.6\%) and high resolution percentage (86.7\%) of documentation PTD , the following resolution practice used by developers in code reviews is \textbf{Documentation improvement} (34 out of 132, 25.8\%). An example of such a case is shown below, where the reviewer recommended adding a comment for the pattern and the developer followed the recommendation of the reviewer and supplemented the missing code comment. It is important for developers to improve the documentation in development process. Better documentation (e.g., source code comments) could make the information easier to access for developers, help the readers better and quicker to grasp the intention of the source code (such as when other developers might want to reuse your code), lighten the burden of maintenance.

\FrameSep=8pt\FrameRule=1pt\begin{framed}\noindent
\textbf{Link:} \url{http://alturl.com/siws8} \\
\textbf{Project:} Qt Base \\
\textbf{Reviewer:} ``I like this pattern, but it deserve a comment:

\textit{// prev is a pointer to the "next" element within the previous node, or to the "firstObserverPtr" if it is the first node.}'' \\
\textbf{Developer:} ``Agreed. Thanks for the comment - added.''
\end{framed}

In addition, 16 (12.1\%) PTD was resolved by \textbf{Testing improvement}, such as the responses of developers in the following examples: ``\textit{Add one test case for the case\footnote{\url{http://alturl.com/37ssk}}}'' and ``\textit{I'll add a simple test to assert the capability...\footnote{\url{http://alturl.com/txfec}}}''. By testing, developers could find bugs or defects in advance, and fix them to guarantee the quality of software. Developers could also evaluate the specific properties of software by testing to make the software more robust and reliable. Also, testing could be helpful to avoid the injection of TD in software systems \cite{10.1145/3387906.3388632}. Closely related to \textbf{Testing improvement}, \textbf{Bug fixing} is mainly used to resolve or correct the defect PTD after finding the defects or bugs. It accounts for 5.3\% (7 out of 132) in our cases.

In the rest (8 out of 132, 6.1\%) PTD, the developers abandoned the code changes and the PTD was thus disappeared with the code. \textbf{Code change abandonment} is considered as a special resolution practice in our study since the PTD is ``resolved'' by discarding the code changes. However, it will lead to extra cost of human resources, time, and budget if the code changes are abandoned in the working progress. Therefore, further investigation on the reasons why developers abandon the code changes is needed.

%
\FrameSep=3pt\FrameRule=0.5pt\begin{framed}
\noindent \textbf{RQ5 Summary:} We have identified five resolution practices to resolve the PTD identified in code reviews. Among them, code refactoring is the main practice, which is used to resolve the majority of the PTD.
\end{framed}

\section{Implications} \label{Implications}

\subsection{Implications for Practitioners}
For project managers, we argue that code review, as a good practice for software quality assurance in software development \cite{wiegers2002peer}, could be used as one of the trustworthy mechanisms to detect PTD in the development process, since our study reveals that reviewers are able to spot PTD that developers are not aware of and let developers resolve PTD at an early stage. Moreover, in our study we find that although the majority of the identified PTD has been resolved by developers, a significant proportion of the identified PTD (19.0\%) was still injected into the software systems without being resolved, which may severely degrade the code quality and impact the maintainability of the systems, especially we observed tha only 16.7\% of requirement PTD was resolved as shown in the results of RQ3. Hence, there is a need for establishing an effective mechanism for PTD management. For the identified PTD, developers should try their best to resolve it before the code is merged into the main repository. If the PTD could not be resolved due to time pressure or other constraints, it should be explicitly recorded and scheduled to be resolved in the future.

For developers, our study finds that most of PTD is identified by reviewers, which implies that developers may write not-quite-right code due to oversight or lack of experience. Therefore, we provide the following suggestions to avoid the occurrence of PTD and resolve PTD:

1) The results of RQ2 shows that code PTD is the majority of PTD, which indicates that developers are more likely to write not-quite-right code. To avoid the occurrence of code PTD, developers should follow closely coding conventions and the best coding practices, such as using meaningful naming for variables and functions in the code, avoiding using magic value, copy/paste activities, and so on. 

2) For documentation PTD, developers should actively write necessary documentation (e.g., code comments), and pay attention to the corresponding documentation when modifying the code to avoid making the documentation outdated and invalid. According to the result of RQ4, documentation PTD costs the less time to resolve compared with other types of PTD, which implies that it is easier to resolve compared to other types of PTD.

3) For test PTD, based on the main reasons that lead to Test PTD in the results of RQ2, developers could consider more about the test coverage of their code and add sufficient test cases in the development process.

4) For design PTD, based on the results of RQ2, poor implementation is the main reason that could result in design PTD. Developers should learn from the suggestions of reviewers and gain knowledge, such as the ideas and insights towards the design issues, which could help them implement the design in a better way. 

5) For requirement PTD, there is still a large proportion of requirement PTD remaining unresolved. Considering the difficulty of resolving requirement PTD, developers could first record them which can be managed and resolved in the future.   

6) For defect PTD, developers should timely resolve it, otherwise it might affect the correctness of the system and incur interest~\cite{li2020interest} when the code contains defect PTD is merged. For the unresolved defect PTD, developers could explicitly record it (e.g., in the commit message or code comment) to inform other developers.

\vspace*{-3pt}
\subsection{Implications for Researchers}
In this work, we focused on PTD, which is an important source of TD. PTD refers to the technical compromises (e.g., sub-optimal implementations) in the development process concerned with the aspects from implementation (i.e., at the code level) to design, and even documentation, requirements, and testing. PTD will evolve into TD when it is injected into the system (e.g., code is merged into the main repository) without being resolved. In our study, we aimed to identify PTD in code review, a process of manually reviewing and verifying code during development. As an exploratory study regarding PTD in code reviews, this work provides researchers with a basic view of PTD by investigating its distribution, classification, and resolution. We believe that the knowledge of PTD could help researchers gain new insights towards the understanding and management of TD. For example, according to the results of RQ3, the resolution proportion of code PTD is very high while the number of code PTD is the highest. This finding indicates that developers show their great emphasis on code PTD, and they resolve PTD with an active attitude. A recent study, that investigated the practitioners' intent for self-fixing TD~\cite{JieTanSelfFix}, came to a similar conclusion that code TD is the main concern compared with other types of TD. Our results of RQ5 reveal that code refactoring is the main practice to resolve PTD. Previous studies on the payment of TD show that code refactoring is also the main practice to pay back TD \cite{10.1145/3383219.3383241, 10.1145/3387906.3388632}. In addition, some TD payment practices are also employed as the resolution practices of PTD, such as \textit{Documentation improvement} corresponds to the payment practice \textit{Updating system documentation} identified in \cite{10.1145/3383219.3383241}, and \textit{Testing improvement} is the same as the payment practice \textit{improve testing} in \cite{10.1145/3387906.3388632}.

Further research could be done towards understanding and managing PTD. Our data was extracted from code review comments. Mining PTD from other sources, such as pull requests together with issue trackers on GitHub, can provide more valuable and comprehensive data for further analysis and comparison. Researchers could pay more attention to the design of automatic tools to identify and classify PTD in review comments, which can help both developers and reviewers better track and deal with PTD in code reviews. In addition, it is also necessary to investigate the reasons for why developers chose not to resolve such PTD before it was injected into the software system, which may help to come up with effective PTD management strategies.

\section{Threats to Validity} \label{Threats to validity}

\textbf{Construct Validity} reflects on the extent of consistency between the operational measures of the study and the RQs \cite{runeson2009guidelines}. A potential threat in this study involves whether the collected review comments were correctly labelled and analyzed by the researchers. In this work, we depended on human activities, which would induce personal bias. To reduce this threat, we labelled the review comments by two researchers independently. Any disagreements were discussed and addressed with a third researcher. Moreover, we conducted a pilot study to make sure that the two researchers achieved a consensus understanding of PTD-related comments. The data extraction process was conducted by one researcher, and the results were reviewed and checked by other two researchers. Another potential threat is whether the dataset is sufficient enough to obtain reasonable conclusions. For RQ1, and RQ2, we randomly selected a representative sample size based on 3\% margin of error and 95\% confidence level~\cite{israel1992dss}. We believe that this measure can partially alleviate this threat. However, there are only 163 PTD-related comments for answering RQ3, RQ4, and RQ5. We admit that the sample size may threaten the validity of our results. Therefore, we conjecture that we could obtain more convincing results by extending this work in a large dataset from more communities and projects, which is also our next step.

\textbf{External Validity} refers to the degree to which our study results and findings can be generalized in other cases (e.g., other projects in OpenStack and Qt, or projects in other communities). We considered the OpenStack and Qt communities since these two communities have made a serious investment in code reviews for several years and are widely used in many studies related to code reviews. As for the selection of projects, we select the most active and major project for each community as our investigated projects. Therefore, we argue that the communities and projects are representative and can increase the generalizability of our study results.

\textbf{Reliability} refers to the replicability of a study for obtaining same or similar results. To improve the reliability, we made a research protocol with detailed procedure, which was discussed and confirmed by all the authors. Besides, all of the empirical steps in our study, including the data mining process, manual data labelling, and manual data extraction and analysis, were conducted and discussed by three authors. Furthermore, the dataset and analysis results of our study have been made publicly available online in order to facilitate other researchers to replicate our study \cite{replication-package}. We believe that these measures can partially alleviate this threat.

\vspace*{-3pt}
\section{Conclusions} \label{Conclusions}
This paper reports on an exploratory study that investigated PTD and its resolution in code reviews. We chose two widely known communities (i.e., OpenStack and Qt) since they have invested heavily in their code reviews for several years and provided reliable peer review activity data. We then randomly collected and analyzed 2,030 review comments from two active projects (i.e., Nova of OpenStack and Qt Base of Qt).

According to our results, PTD is not prevalently identified in code reviews, and when identified, code and documentation PTD are the most frequently identified PTD among six types of PTD (i.e., design, defect, documentation, requirement, test, and code PTD). This finding indicates that developers should care more about their code quality along with the corresponding documentation when implementing and modifying the code. We also found that around 81.0\% of the PTD identified in code reviews has been resolved by developers and 78.0\% of the resolved PTD was resolved by developers within a week. We conjectured that this is mainly because the context is reviewer-centric code review process (e.g., reviewers sometimes provide constructive recommendations to help developers resolve PTD). In addition, it usually takes developers less time to resolve documentation PTD, implying that documentation PTD may be easier to resolve compared with other types of PTD. Although the majority of the identified PTD has been resolved by developers, a significant proportion of identified PTD (19.0\%) was still injected into the systems without being resolved, especially for requirement PTD (i.e., only 16.7\% of requirement PTD was resolved). This finding suggests a need for project managers to establish an effective mechanism for PTD management in their projects to prevent the negative impact when PTD evolves into TD. Moreover, code refactoring is the main practice used to resolve PTD identified in code reviews.  

Given the importance of PTD in development, we plan to extend this work by studying PTD in code reviews in a large dataset from different communities and projects to further understand and analyze PTD identified and resolved in code reviews, the reasons of not resolving PTD, as well as the human factors in this process.

\vspace*{-3pt}
\begin{acks}
This work is funded by NSFC with No. 62172311, Hubei Provincial Natural Science Foundation of China with No. 2021CFB577, and the Special Fund of Hubei Luojia Laboratory. Amjed Tahir is supported by a NZ SfTI National Science Challenge Grant (MAUX2004).
\end{acks}

\balance
\bibliographystyle{ACM-Reference-Format}
\bibliography{ref}


\begin{thebibliography}{38}


\ifx \showCODEN    \undefined \def \showCODEN     #1{\unskip}     \fi
\ifx \showDOI      \undefined \def \showDOI       #1{#1}\fi
\ifx \showISBNx    \undefined \def \showISBNx     #1{\unskip}     \fi
\ifx \showISBNxiii \undefined \def \showISBNxiii  #1{\unskip}     \fi
\ifx \showISSN     \undefined \def \showISSN      #1{\unskip}     \fi
\ifx \showLCCN     \undefined \def \showLCCN      #1{\unskip}     \fi
\ifx \shownote     \undefined \def \shownote      #1{#1}          \fi
\ifx \showarticletitle \undefined \def \showarticletitle #1{#1}   \fi
\ifx \showURL      \undefined \def \showURL       {\relax}        \fi
\providecommand\bibfield[2]{#2}
\providecommand\bibinfo[2]{#2}
\providecommand\natexlab[1]{#1}
\providecommand\showeprint[2][]{arXiv:#2}

\bibitem[\protect\citeauthoryear{Basili, Caldiera, and Rombach}{Basili
  et~al\mbox{.}}{1994}]%
        {caldiera1994goal}
\bibfield{author}{\bibinfo{person}{Victor~R. Basili},
  \bibinfo{person}{Gianluigi Caldiera}, {and} \bibinfo{person}{H.~D. Rombach}.}
  \bibinfo{year}{1994}\natexlab{}.
\newblock \showarticletitle{The Goal Question Metric Approach}.
\newblock \bibinfo{journal}{\emph{Encyclopedia of Software Engineering}}
  (\bibinfo{year}{1994}), \bibinfo{pages}{528--532}.
\newblock


\bibitem[\protect\citeauthoryear{Baysal, Kononenko, Holmes, and Godfrey}{Baysal
  et~al\mbox{.}}{2013}]%
        {6671287}
\bibfield{author}{\bibinfo{person}{Olga Baysal}, \bibinfo{person}{Oleksii
  Kononenko}, \bibinfo{person}{Reid Holmes}, {and} \bibinfo{person}{Michael~W.
  Godfrey}.} \bibinfo{year}{2013}\natexlab{}.
\newblock \showarticletitle{The Influence of Non-technical Factors on Code
  Review}. In \bibinfo{booktitle}{\emph{Proceedings of the 20th Working
  Conference on Reverse Engineering (WCRE)}}. \bibinfo{publisher}{IEEE},
  \bibinfo{pages}{122--131}.
\newblock


\bibitem[\protect\citeauthoryear{Cohen}{Cohen}{1960}]%
        {cohen1960coefficient}
\bibfield{author}{\bibinfo{person}{Jacob Cohen}.}
  \bibinfo{year}{1960}\natexlab{}.
\newblock \showarticletitle{A Coefficient of Agreement for Nominal Scales}.
\newblock \bibinfo{journal}{\emph{Educational and Psychological Measurement}}
  \bibinfo{volume}{20}, \bibinfo{number}{1} (\bibinfo{year}{1960}),
  \bibinfo{pages}{37--46}.
\newblock


\bibitem[\protect\citeauthoryear{Cunningham}{Cunningham}{1992}]%
        {CunninghamMS}
\bibfield{author}{\bibinfo{person}{Ward Cunningham}.}
  \bibinfo{year}{1992}\natexlab{}.
\newblock \showarticletitle{The WyCash Portfolio Management System}.
\newblock \bibinfo{journal}{\emph{ACM SIGPLAN OOPS Messenger}}
  \bibinfo{volume}{4}, \bibinfo{number}{2} (\bibinfo{year}{1992}),
  \bibinfo{pages}{29–--30}.
\newblock


\bibitem[\protect\citeauthoryear{de~Mello, Oliveira, and Garcia}{de~Mello
  et~al\mbox{.}}{2017}]%
        {8170086}
\bibfield{author}{\bibinfo{person}{Rafael~Maiani de Mello},
  \bibinfo{person}{Roberto Oliveira}, {and} \bibinfo{person}{Alessandro
  Garcia}.} \bibinfo{year}{2017}\natexlab{}.
\newblock \showarticletitle{On the Influence of Human Factors for Identifying
  Code Smells: A Multi-Trial Empirical Study}. In
  \bibinfo{booktitle}{\emph{Proceedings of the 11th ACM/IEEE International
  Symposium on Empirical Software Engineering and Measurement (ESEM)}}.
  \bibinfo{publisher}{IEEE}, \bibinfo{pages}{68--77}.
\newblock


\bibitem[\protect\citeauthoryear{de~Toledo, Martini, Przybyszewska, and
  Sj{\o}berg}{de~Toledo et~al\mbox{.}}{2019}]%
        {de2019architectural}
\bibfield{author}{\bibinfo{person}{Saulo~Soares de Toledo},
  \bibinfo{person}{Antonio Martini}, \bibinfo{person}{Agata Przybyszewska},
  {and} \bibinfo{person}{Dag~IK Sj{\o}berg}.} \bibinfo{year}{2019}\natexlab{}.
\newblock \showarticletitle{Architectural Technical Debt in Microservices: A
  Case Study in a Large Company}. In \bibinfo{booktitle}{\emph{Proceedings of
  the 2rd IEEE/ACM International Conference on Technical Debt (TechDebt)}}.
  IEEE, \bibinfo{pages}{78--87}.
\newblock


\bibitem[\protect\citeauthoryear{Fowler}{Fowler}{2009}]%
        {flower}
\bibfield{author}{\bibinfo{person}{Martin Fowler}.}
  \bibinfo{year}{2009}\natexlab{}.
\newblock \showarticletitle{Technical Debt Quadrant}.
\newblock
\urldef\tempurl%
\url{https://martinfowler.com/bliki/TechnicalDebtQuadrant.html}
\showURL{%
\tempurl}


\bibitem[\protect\citeauthoryear{Fowler}{Fowler}{2018}]%
        {fowler2018refactoring}
\bibfield{author}{\bibinfo{person}{Martin Fowler}.}
  \bibinfo{year}{2018}\natexlab{}.
\newblock \bibinfo{booktitle}{\emph{Refactoring: Improving the Design of
  Existing Code}}.
\newblock \bibinfo{publisher}{Addison-Wesley Professional}.
\newblock


\bibitem[\protect\citeauthoryear{Freire, Rios, Gutierrez, Torres, Mendon\c{c}a,
  Izurieta, Seaman, and Sp\'{\i}nola}{Freire et~al\mbox{.}}{2020}]%
        {10.1145/3383219.3383241}
\bibfield{author}{\bibinfo{person}{S\'{a}vio Freire}, \bibinfo{person}{Nicolli
  Rios}, \bibinfo{person}{Boris Gutierrez}, \bibinfo{person}{Dar\'{\i}o
  Torres}, \bibinfo{person}{Manoel Mendon\c{c}a}, \bibinfo{person}{Clemente
  Izurieta}, \bibinfo{person}{Carolyn Seaman}, {and}
  \bibinfo{person}{Rodrigo~O. Sp\'{\i}nola}.} \bibinfo{year}{2020}\natexlab{}.
\newblock \showarticletitle{Surveying Software Practitioners on Technical Debt
  Payment Practices and Reasons for Not Paying off Debt Items}. In
  \bibinfo{booktitle}{\emph{Proceedings of the 24th Evaluation and Assessment
  in Software Engineering (EASE)}}. \bibinfo{publisher}{ACM},
  \bibinfo{pages}{210–219}.
\newblock


\bibitem[\protect\citeauthoryear{Fu, Liang, Rasheed, Li, Tahir, and Han}{Fu
  et~al\mbox{.}}{2022}]%
        {replication-package}
\bibfield{author}{\bibinfo{person}{Liming Fu}, \bibinfo{person}{Peng Liang},
  \bibinfo{person}{Zeeshan Rasheed}, \bibinfo{person}{Zengyang Li},
  \bibinfo{person}{Amjed Tahir}, {and} \bibinfo{person}{Xiaofeng Han}.}
  \bibinfo{year}{2022}\natexlab{}.
\newblock \showarticletitle{Replication Package for the Paper: ``Understanding
  Potential Technical Debt and Its Resolution in Code Reviews : An Exploratory
  Study of the OpenStack and Qt Communities''}.
\newblock
\urldef\tempurl%
\url{https://doi.org/10.5281/zenodo.6513444}
\showURL{%
\tempurl}


\bibitem[\protect\citeauthoryear{Glaser}{Glaser}{1965}]%
        {glaser1965constant}
\bibfield{author}{\bibinfo{person}{Barney~G Glaser}.}
  \bibinfo{year}{1965}\natexlab{}.
\newblock \showarticletitle{The Constant Comparative Method of Qualitative
  Analysis}.
\newblock \bibinfo{journal}{\emph{Social Problems}} \bibinfo{volume}{12},
  \bibinfo{number}{4} (\bibinfo{year}{1965}), \bibinfo{pages}{436--445}.
\newblock


\bibitem[\protect\citeauthoryear{Hamasaki, Kula, Yoshida, Cruz, Fujiwara, and
  Iida}{Hamasaki et~al\mbox{.}}{2013}]%
        {6624003}
\bibfield{author}{\bibinfo{person}{Kazuki Hamasaki},
  \bibinfo{person}{Raula~Gaikovina Kula}, \bibinfo{person}{Norihiro Yoshida},
  \bibinfo{person}{A.~E.~Camargo Cruz}, \bibinfo{person}{Kenji Fujiwara}, {and}
  \bibinfo{person}{Hajimu Iida}.} \bibinfo{year}{2013}\natexlab{}.
\newblock \showarticletitle{Who Does What during a Code Review? Datasets of OSS
  Peer Review Repositories}. In \bibinfo{booktitle}{\emph{Proceedings of the
  10th Working Conference on Mining Software Repositories (MSR)}}.
  \bibinfo{publisher}{ACM}, \bibinfo{pages}{49--52}.
\newblock


\bibitem[\protect\citeauthoryear{Han, Tahir, Liang, Counsell, Blincoe, Li, and
  Luo}{Han et~al\mbox{.}}{2022}]%
        {han2022code}
\bibfield{author}{\bibinfo{person}{Xiaofeng Han}, \bibinfo{person}{Amjed
  Tahir}, \bibinfo{person}{Peng Liang}, \bibinfo{person}{Steve Counsell},
  \bibinfo{person}{Kelly Blincoe}, \bibinfo{person}{Bing Li}, {and}
  \bibinfo{person}{Yajing Luo}.} \bibinfo{year}{2022}\natexlab{}.
\newblock \showarticletitle{Code Smells Detection via Modern Code Review: A
  Study of the OpenStack and Qt Communities}.
\newblock \bibinfo{journal}{\emph{Empirical Software Engineering}}
  (\bibinfo{year}{2022}).
\newblock


\bibitem[\protect\citeauthoryear{Han, Tahir, Liang, Counsell, and Luo}{Han
  et~al\mbox{.}}{2021}]%
        {hanCodeSmell}
\bibfield{author}{\bibinfo{person}{Xiaofeng Han}, \bibinfo{person}{Amjed
  Tahir}, \bibinfo{person}{Peng Liang}, \bibinfo{person}{Steve Counsell}, {and}
  \bibinfo{person}{Yajing Luo}.} \bibinfo{year}{2021}\natexlab{}.
\newblock \showarticletitle{Understanding Code Smell Detection via Code Review:
  A Study of the OpenStack Community}. In
  \bibinfo{booktitle}{\emph{Procceedings of the 29th IEEE/ACM International
  Conference on Program Comprehension (ICPC)}}. \bibinfo{publisher}{IEEE},
  \bibinfo{pages}{323--334}.
\newblock


\bibitem[\protect\citeauthoryear{Hirao, McIntosh, Ihara, and Matsumoto}{Hirao
  et~al\mbox{.}}{2022}]%
        {HiraoCode}
\bibfield{author}{\bibinfo{person}{Toshiki Hirao}, \bibinfo{person}{Shane
  McIntosh}, \bibinfo{person}{Akinori Ihara}, {and} \bibinfo{person}{Kenichi
  Matsumoto}.} \bibinfo{year}{2022}\natexlab{}.
\newblock \showarticletitle{Code Reviews with Divergent Review Scores: An
  Empirical Study of the OpenStack and Qt Communities}.
\newblock \bibinfo{journal}{\emph{IEEE Transactions on Software Engineering}}
  \bibinfo{volume}{48}, \bibinfo{number}{1} (\bibinfo{year}{2022}),
  \bibinfo{pages}{69--81}.
\newblock


\bibitem[\protect\citeauthoryear{Israel}{Israel}{1992}]%
        {israel1992dss}
\bibfield{author}{\bibinfo{person}{Glenn~D. Israel}.}
  \bibinfo{year}{1992}\natexlab{}.
\newblock \bibinfo{booktitle}{\emph{Determining Sample Size}}.
\newblock \bibinfo{type}{Fact Sheet} PEOD-6. \bibinfo{institution}{Florida
  Cooperative Extension Service, Institute of Food and Agricultural Sciences,
  University of Florida}, \bibinfo{address}{Florida, U.S.A}.
\newblock


\bibitem[\protect\citeauthoryear{Kashiwa, Nishikawa, Kamei, Kondo, Shihab,
  Sato, and Ubayashi}{Kashiwa et~al\mbox{.}}{2022}]%
        {KASHIWA2022106855}
\bibfield{author}{\bibinfo{person}{Yutaro Kashiwa}, \bibinfo{person}{Ryoma
  Nishikawa}, \bibinfo{person}{Yasutaka Kamei}, \bibinfo{person}{Masanari
  Kondo}, \bibinfo{person}{Emad Shihab}, \bibinfo{person}{Ryosuke Sato}, {and}
  \bibinfo{person}{Naoyasu Ubayashi}.} \bibinfo{year}{2022}\natexlab{}.
\newblock \showarticletitle{An Empirical Study on Self-Admitted Technical Debt
  in Modern Code Review}.
\newblock \bibinfo{journal}{\emph{Information and Software Technology}}
  \bibinfo{volume}{146} (\bibinfo{year}{2022}), \bibinfo{pages}{106855}.
\newblock


\bibitem[\protect\citeauthoryear{Kipyegen and Korir}{Kipyegen and
  Korir}{2013}]%
        {isd}
\bibfield{author}{\bibinfo{person}{Noela~J. Kipyegen} {and}
  \bibinfo{person}{William P.~K. Korir}.} \bibinfo{year}{2013}\natexlab{}.
\newblock \showarticletitle{Importance of Software Documentation}.
\newblock \bibinfo{journal}{\emph{International Journal of Computer Science
  Issues (IJCSI)}} \bibinfo{volume}{10}, \bibinfo{number}{5}
  (\bibinfo{year}{2013}), \bibinfo{pages}{223--228}.
\newblock


\bibitem[\protect\citeauthoryear{Kruchten, Nord, and Ozkaya}{Kruchten
  et~al\mbox{.}}{2012}]%
        {kruchten2012technical}
\bibfield{author}{\bibinfo{person}{Philippe Kruchten},
  \bibinfo{person}{Robert~L Nord}, {and} \bibinfo{person}{Ipek Ozkaya}.}
  \bibinfo{year}{2012}\natexlab{}.
\newblock \showarticletitle{Technical Debt: From Metaphor to Theory and
  Practice}.
\newblock \bibinfo{journal}{\emph{IEEE Software}} \bibinfo{volume}{29},
  \bibinfo{number}{6} (\bibinfo{year}{2012}), \bibinfo{pages}{18--21}.
\newblock


\bibitem[\protect\citeauthoryear{Lefever, Cai, Cervantes, Kazman, and
  Fang}{Lefever et~al\mbox{.}}{2021}]%
        {lefever2021lack}
\bibfield{author}{\bibinfo{person}{Jason Lefever}, \bibinfo{person}{Yuanfang
  Cai}, \bibinfo{person}{Humberto Cervantes}, \bibinfo{person}{Rick Kazman},
  {and} \bibinfo{person}{Hongzhou Fang}.} \bibinfo{year}{2021}\natexlab{}.
\newblock \showarticletitle{On the Lack of Consensus Among Technical Debt
  Detection Tools}. In \bibinfo{booktitle}{\emph{Proceedings of the 43rd
  IEEE/ACM International Conference on Software Engineering: Software
  Engineering in Practice (ICSE-SEIP)}}. IEEE, \bibinfo{pages}{121--130}.
\newblock


\bibitem[\protect\citeauthoryear{Li, Soliman, and Avgeriou}{Li
  et~al\mbox{.}}{2020a}]%
        {9226330}
\bibfield{author}{\bibinfo{person}{Yikun Li}, \bibinfo{person}{Mohamed
  Soliman}, {and} \bibinfo{person}{Paris Avgeriou}.}
  \bibinfo{year}{2020}\natexlab{a}.
\newblock \showarticletitle{Identification and Remediation of Self-Admitted
  Technical Debt in Issue Trackers}. In \bibinfo{booktitle}{\emph{Proceedings
  of the 46th Euromicro Conference on Software Engineering and Advanced
  Applications (SEAA)}}. \bibinfo{publisher}{IEEE}, \bibinfo{pages}{495--503}.
\newblock


\bibitem[\protect\citeauthoryear{Li, Avgeriou, and Liang}{Li
  et~al\mbox{.}}{2015}]%
        {li2015systematic}
\bibfield{author}{\bibinfo{person}{Zengyang Li}, \bibinfo{person}{Paris
  Avgeriou}, {and} \bibinfo{person}{Peng Liang}.}
  \bibinfo{year}{2015}\natexlab{}.
\newblock \showarticletitle{A systematic mapping study on technical debt and
  its management}.
\newblock \bibinfo{journal}{\emph{Journal of Systems and Software}}
  \bibinfo{volume}{101} (\bibinfo{year}{2015}), \bibinfo{pages}{193--220}.
\newblock


\bibitem[\protect\citeauthoryear{Li, Yu, Liang, Mo, and Yang}{Li
  et~al\mbox{.}}{2020b}]%
        {li2020interest}
\bibfield{author}{\bibinfo{person}{Zengyang Li}, \bibinfo{person}{Qinyi Yu},
  \bibinfo{person}{Peng Liang}, \bibinfo{person}{Ran Mo}, {and}
  \bibinfo{person}{Chen Yang}.} \bibinfo{year}{2020}\natexlab{b}.
\newblock \showarticletitle{Interest of Defect Technical Debt: An Exploratory
  Study on Apache Projects}. In \bibinfo{booktitle}{\emph{Proceedings fo the
  36th IEEE International Conference on Software Maintenance and Evolution
  (ICSME)}}. \bibinfo{publisher}{IEEE}, \bibinfo{pages}{629--639}.
\newblock


\bibitem[\protect\citeauthoryear{Liu, Huang, Xia, Shihab, Lo, and Li}{Liu
  et~al\mbox{.}}{2020}]%
        {tdindl}
\bibfield{author}{\bibinfo{person}{Jiakun Liu}, \bibinfo{person}{Qiao Huang},
  \bibinfo{person}{Xin Xia}, \bibinfo{person}{Emad Shihab},
  \bibinfo{person}{David Lo}, {and} \bibinfo{person}{Shanping Li}.}
  \bibinfo{year}{2020}\natexlab{}.
\newblock \showarticletitle{Is Using Deep Learning Frameworks Free?
  Characterizing Technical Debt in Deep Learning Frameworks}. In
  \bibinfo{booktitle}{\emph{Proceedings of the 42nd ACM/IEEE International
  Conference on Software Engineering: Software Engineering in Society
  (ICSE-SEIS)}}. \bibinfo{publisher}{ACM}, \bibinfo{pages}{1–10}.
\newblock


\bibitem[\protect\citeauthoryear{Maldonado and Shihab}{Maldonado and
  Shihab}{2015}]%
        {MaldonadoDetect}
\bibfield{author}{\bibinfo{person}{Everton da~S. Maldonado} {and}
  \bibinfo{person}{Emad Shihab}.} \bibinfo{year}{2015}\natexlab{}.
\newblock \showarticletitle{Detecting and Quantifying Different Types of
  Self-Admitted Technical Debt}. In \bibinfo{booktitle}{\emph{Proceedings of
  the 7th IEEE International Workshop on Managing Technical Debt (MTD)}}.
  \bibinfo{publisher}{IEEE}, \bibinfo{pages}{9--15}.
\newblock


\bibitem[\protect\citeauthoryear{McIntosh, Kamei, Adams, and Hassan}{McIntosh
  et~al\mbox{.}}{2016}]%
        {mcintosh2016empirical}
\bibfield{author}{\bibinfo{person}{Shane McIntosh}, \bibinfo{person}{Yasutaka
  Kamei}, \bibinfo{person}{Bram Adams}, {and} \bibinfo{person}{Ahmed~E
  Hassan}.} \bibinfo{year}{2016}\natexlab{}.
\newblock \showarticletitle{An Empirical Study of the Impact of Modern Code
  Review Practices on Software Quality}.
\newblock \bibinfo{journal}{\emph{Empirical Software Engineering}}
  \bibinfo{volume}{21}, \bibinfo{number}{5} (\bibinfo{year}{2016}),
  \bibinfo{pages}{2146--2189}.
\newblock


\bibitem[\protect\citeauthoryear{Nanthaamornphong and
  Chaisutanon}{Nanthaamornphong and Chaisutanon}{2016}]%
        {nanthaamornphong2016empirical}
\bibfield{author}{\bibinfo{person}{Aziz Nanthaamornphong} {and}
  \bibinfo{person}{Apatta Chaisutanon}.} \bibinfo{year}{2016}\natexlab{}.
\newblock \showarticletitle{Empirical Evaluation of Code Smells in Open Source
  Projects: Preliminary Results}. In \bibinfo{booktitle}{\emph{Proceedings of
  the 1st International Workshop on Software Refactoring (IWoR)}}.
  \bibinfo{publisher}{ACM}, \bibinfo{pages}{5--8}.
\newblock


\bibitem[\protect\citeauthoryear{P\'{e}rez, Castellanos, Correal, Rios, Freire,
  Sp\'{\i}nola, and Seaman}{P\'{e}rez et~al\mbox{.}}{2020}]%
        {10.1145/3387906.3388632}
\bibfield{author}{\bibinfo{person}{Boris P\'{e}rez}, \bibinfo{person}{Camilo
  Castellanos}, \bibinfo{person}{Dar\'{\i}o Correal}, \bibinfo{person}{Nicolli
  Rios}, \bibinfo{person}{S\'{a}vio Freire}, \bibinfo{person}{Rodrigo
  Sp\'{\i}nola}, {and} \bibinfo{person}{Carolyn Seaman}.}
  \bibinfo{year}{2020}\natexlab{}.
\newblock \showarticletitle{What Are the Practices Used by Software
  Practitioners on Technical Debt Payment: Results from an International Family
  of Surveys}. In \bibinfo{booktitle}{\emph{Proceedings of the 3rd
  International Conference on Technical Debt (TechDebt)}}.
  \bibinfo{publisher}{ACM}, \bibinfo{pages}{103–112}.
\newblock


\bibitem[\protect\citeauthoryear{Potdar and Shihab}{Potdar and Shihab}{2014}]%
        {potdar2014exploratory}
\bibfield{author}{\bibinfo{person}{Aniket Potdar} {and} \bibinfo{person}{Emad
  Shihab}.} \bibinfo{year}{2014}\natexlab{}.
\newblock \showarticletitle{An Exploratory Study on Self-Admitted Technical
  Debt}. In \bibinfo{booktitle}{\emph{Proceedings of the 30th IEEE
  International Conference on Software Maintenance and Evolution (ICSME)}}.
  IEEE, \bibinfo{pages}{91--100}.
\newblock


\bibitem[\protect\citeauthoryear{Runeson and H{\"o}st}{Runeson and
  H{\"o}st}{2009}]%
        {runeson2009guidelines}
\bibfield{author}{\bibinfo{person}{Per Runeson} {and} \bibinfo{person}{Martin
  H{\"o}st}.} \bibinfo{year}{2009}\natexlab{}.
\newblock \showarticletitle{Guidelines for Conducting and Reporting Case Study
  Research in Software Engineering}.
\newblock \bibinfo{journal}{\emph{Empirical Software Engineering}}
  \bibinfo{volume}{14}, \bibinfo{number}{2} (\bibinfo{year}{2009}),
  \bibinfo{pages}{131--164}.
\newblock


\bibitem[\protect\citeauthoryear{Schmid}{Schmid}{2013}]%
        {6608681}
\bibfield{author}{\bibinfo{person}{Klaus Schmid}.}
  \bibinfo{year}{2013}\natexlab{}.
\newblock \showarticletitle{On the Limits of the Technical Debt Metaphor Some
  Guidance on Going Beyond}. In \bibinfo{booktitle}{\emph{Proceedings of the
  4th International Workshop on Managing Technical Debt (MTD)}}.
  \bibinfo{publisher}{IEEE}, \bibinfo{pages}{63--66}.
\newblock


\bibitem[\protect\citeauthoryear{Silva, Valente, and Terra}{Silva
  et~al\mbox{.}}{2016}]%
        {silva2016does}
\bibfield{author}{\bibinfo{person}{Marcelino Campos~Oliveira Silva},
  \bibinfo{person}{Marco~Tulio Valente}, {and} \bibinfo{person}{Ricardo
  Terra}.} \bibinfo{year}{2016}\natexlab{}.
\newblock \showarticletitle{Does Technical Debt Lead to the Rejection of Pull
  Requests?}. In \bibinfo{booktitle}{\emph{Proceedings of the 12th Brazilian
  Symposium on Information Systems (SBSI)}}. \bibinfo{publisher}{ACM},
  \bibinfo{pages}{248--254}.
\newblock


\bibitem[\protect\citeauthoryear{Tan, Feitosa, and Avgeriou}{Tan
  et~al\mbox{.}}{2021}]%
        {JieTanSelfFix}
\bibfield{author}{\bibinfo{person}{Jie Tan}, \bibinfo{person}{Daniel Feitosa},
  {and} \bibinfo{person}{Paris Avgeriou}.} \bibinfo{year}{2021}\natexlab{}.
\newblock \showarticletitle{Do Practitioners Intentionally Self-Fix Technical
  Debt and Why?}. In \bibinfo{booktitle}{\emph{Proceedings of the 37th IEEE
  International Conference on Software Maintenance and Evolution (ICSME)}}.
  \bibinfo{publisher}{IEEE}, \bibinfo{pages}{251--262}.
\newblock


\bibitem[\protect\citeauthoryear{Wang, Xiao, Thongtanunam, Kula, and
  Matsumoto}{Wang et~al\mbox{.}}{2021}]%
        {wang2021understanding}
\bibfield{author}{\bibinfo{person}{Dong Wang}, \bibinfo{person}{Tao Xiao},
  \bibinfo{person}{Patanamon Thongtanunam}, \bibinfo{person}{Raula~Gaikovina
  Kula}, {and} \bibinfo{person}{Kenichi Matsumoto}.}
  \bibinfo{year}{2021}\natexlab{}.
\newblock \showarticletitle{Understanding Shared Links and Their Intentions to
  Meet Information Needs in Modern Code Review}.
\newblock \bibinfo{journal}{\emph{Empirical Software Engineering}}
  \bibinfo{volume}{26}, \bibinfo{number}{5} (\bibinfo{year}{2021}),
  \bibinfo{pages}{1--32}.
\newblock


\bibitem[\protect\citeauthoryear{Wiegers}{Wiegers}{2002}]%
        {wiegers2002peer}
\bibfield{author}{\bibinfo{person}{Karl~Eugene Wiegers}.}
  \bibinfo{year}{2002}\natexlab{}.
\newblock \bibinfo{booktitle}{\emph{Peer Reviews in Software: A Practical
  Guide}}.
\newblock \bibinfo{publisher}{Addison-Wesley Boston}.
\newblock


\bibitem[\protect\citeauthoryear{Zanaty, Hirao, McIntosh, Ihara, and
  Matsumoto}{Zanaty et~al\mbox{.}}{2018}]%
        {designDiscussion}
\bibfield{author}{\bibinfo{person}{Farida~El Zanaty}, \bibinfo{person}{Toshiki
  Hirao}, \bibinfo{person}{Shane McIntosh}, \bibinfo{person}{Akinori Ihara},
  {and} \bibinfo{person}{Kenichi Matsumoto}.} \bibinfo{year}{2018}\natexlab{}.
\newblock \showarticletitle{An Empirical Study of Design Discussions in Code
  Review}. In \bibinfo{booktitle}{\emph{Proceedings of the 12th ACM/IEEE
  International Symposium on Empirical Software Engineering and Measurement
  (ESEM)}}. \bibinfo{publisher}{ACM}, Article \bibinfo{articleno}{11}.
\newblock


\bibitem[\protect\citeauthoryear{Zazworka, Shaw, Shull, and Seaman}{Zazworka
  et~al\mbox{.}}{2011}]%
        {DesignDebtZazworka}
\bibfield{author}{\bibinfo{person}{Nico Zazworka}, \bibinfo{person}{Michele~A.
  Shaw}, \bibinfo{person}{Forrest Shull}, {and} \bibinfo{person}{Carolyn
  Seaman}.} \bibinfo{year}{2011}\natexlab{}.
\newblock \showarticletitle{Investigating the Impact of Design Debt on Software
  Quality}. In \bibinfo{booktitle}{\emph{Proceedings of the 2nd Workshop on
  Managing Technical Debt (MTD)}}. \bibinfo{publisher}{ACM},
  \bibinfo{pages}{17--–23}.
\newblock


\bibitem[\protect\citeauthoryear{Zazworka, Sp{\'\i}nola, Vetro', Shull, and
  Seaman}{Zazworka et~al\mbox{.}}{2013}]%
        {zazworka2013case}
\bibfield{author}{\bibinfo{person}{Nico Zazworka}, \bibinfo{person}{Rodrigo~O
  Sp{\'\i}nola}, \bibinfo{person}{Antonio Vetro'}, \bibinfo{person}{Forrest
  Shull}, {and} \bibinfo{person}{Carolyn Seaman}.}
  \bibinfo{year}{2013}\natexlab{}.
\newblock \showarticletitle{A Case Study on Effectively Identifying Technical
  Debt}. In \bibinfo{booktitle}{\emph{Proceedings of the 17th International
  Conference on Evaluation and Assessment in Software Engineering (EASE)}}.
  \bibinfo{publisher}{ACM}, \bibinfo{pages}{42--47}.
\newblock


\end{thebibliography}

\end{document}